# A Generalization of Shell Theorem.
# Electric Potential of Charged Spheres and Charged Vesicles Surrounded by Electrolyte

István P. Sugár

Department of Neurology, Icahn School of Medicine at Mount Sinai, New York, NY 10029

**Abstract**
By using the Debye screened potential a generalized version of the Shell Theorem is developed and analytical equations are derived to calculate i) the potential of a charged sphere surrounded by electrolyte, ii) the potential of two concentric charged spheres surrounded by electrolyte, and iii) the membrane potential of a charged lipid vesicle surrounded by electrolyte with high ion concentration. By numerical integration the potential of a lipid vesicle is calculated at any electrolyte concentration.

1. Introduction

The headgroups of mebrane lipids have either single charge (e.g. tetraether lipids [1,2]) or electric dipole (e.g. phospholipids [1,3]). Theoretical models of lipid membranes usually focus on short range (Van der Waals) lateral interactions between nearest neighbor lipids and ignore the long range charge-charge interactions [3,4]. This is because in the case of long range interactions one has to consider the entire system rather than the interactions between the nearest neighbor lipids. In physics the Shell Theorem deals with a similar problem, determining the electric potential within and around a charged sphere (where there is vacuum inside and around the charged sphere of radius R). According to the Shell theorem [5] the electric potential inside the charged sphere is constant, i.e.:

$$V(r) = \frac{k_e Q}{R} \tag{1}$$

where r<R is the distance from the center of the charged sphere and Q is the sum of the charges spreaded homogeneously on the sphere surface and $k_e (= 9 \cdot 10^9 Nm^2 C^{-2})$ is the Coulomb's constant. However if r>R then the potential decreases with increasing distance, i.e.:

$$V(r) = \frac{k_e Q}{r}. \tag{2}$$

In this paper step by step we generalize the Shell Theorem to get closer to the conditions of charged vesicles. First we consider a charged sphere which, like a charged vesicle, is filled and surrounded by electrolyte. Second, to imitate the inner and outer surface of the charged vesicle membrane, we consider two concentric charged spheres with electrolyte all around. Third, since the inside of the lipid membrane is hydrophobic (with dielectric constant $\varepsilon_m = 2$), we consider the two concentric charged spheres with electrolyte all around except between the region of the two spheres. Generalizing the Shell Theorem we found analytical solution for the potential in the first and second cases. In the third case, however, we got analytical solution only if the ion concentration of the electrolyte is so high that the Debye length is much shorter than the membrane thickness. In the cases of longer Debye lengths, i.e. at lower electrolyte concentrations, we provide numerical solutions for the potential.

## 2. Model
### 2.1 Potential around a charged sphere surrounded by electrolyte

According to Debye screening in electrolyte the potential produced at a distance $r$ from an external point charge, $q$ is [6,7]:

$$V(r) = \frac{k_e q}{\varepsilon r} e^{-r/\lambda_D} \tag{3}$$

where $\varepsilon$ and $\lambda_D$ is the relative static permittivity and Debye length, respectively of the electrolyte. In our model (shown in Figure 1) a charged sphere of radius $R1$ is surrounded by electrolyte.

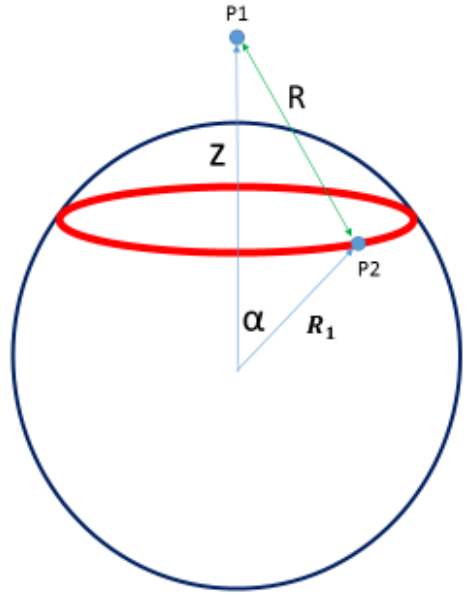

**Figure 1.** *Charged sphere*
Black circle represents the charged sphere of radius $R_1$. The potential is calculated at point P1.

Its distance from the center of the sphere is Z. Red ring represent charges on the charged sphere. Their distance from point P1 is R. $\alpha$ is the angle between vector $Z$ and a vector pointing from the center of the sphere to any of the point (P2) of the red ring.

On the sphere point charges are homogeneously distributed with charge density $\rho_1$. We would like to calculate the potential at point P1, located at a distance $Z > R_1$ from the center of the charged sphere (see Figure 1). First we calculate the potential induced by the charges located along the red ring shown in Figure 1:

$$V(\alpha, Z) \cdot d\alpha = \frac{k_e \cdot \rho_1 \cdot 2 \cdot R_1 \cdot \sin(\alpha) \cdot \pi \cdot R_1 \cdot d\alpha}{\varepsilon R(\alpha, Z, R_1)} e^{-R(\alpha, Z, R_1)/\lambda_D} \tag{4}$$

where $2 \cdot R_1 \cdot \sin(\alpha) \cdot \pi \cdot R_1 \cdot d\alpha$ is the surface area of the red ring and $R(\alpha, Z, R_1)$ is the distance between point P1 and any of the point charges located on the red ring:

$$R(\alpha, Z, R_1) = \sqrt{(R_1 \cdot \sin(\alpha))^2 + (Z - R_1 \cdot \cos(\alpha))^2} = \sqrt{R_1^2 + Z^2 - 2 \cdot Z \cdot R_1 \cdot \cos(\alpha)} \tag{5}$$

The potential induced by the entire charged sphere at point P1 is:

$$V(Z) = \int_0^\pi V(\alpha, Z) d\alpha = \int_0^\pi \frac{k_e \cdot \rho_1 \cdot 2 \cdot R_1 \cdot \sin(\alpha) \cdot \pi \cdot R_1 \cdot d\alpha}{\varepsilon R(\alpha, Z, R_1)} e^{-R(\alpha, Z, R_1)/\lambda_D} \tag{6}$$

After the integration (see Appendix 1) the potential at $Z > R_1$ is

$$V(Z) = \frac{k_e \cdot Q_1 \cdot \lambda_D}{\varepsilon \cdot Z \cdot R_1} \cdot e^{-\frac{Z}{\lambda_D}} \cdot \sinh\left(\frac{R_1}{\lambda_D}\right) \tag{7}$$

where $Q_1 = 4R_1^2\pi \cdot \rho_1$ is the total charge of the sphere, and the potential at $Z < R_1$ is

$$V(Z) = \frac{k_e \cdot Q_1 \cdot \lambda_D}{\varepsilon \cdot Z \cdot R_1} \cdot e^{-\frac{R_1}{\lambda_D}} \cdot \sinh\left(\frac{Z}{\lambda_D}\right) \tag{8}$$

### 2.2 Potential around two concentric charged spheres surrounded by electrolyte

Let us consider two concentric charged spheres, sphere 1 and 2, with radius $R_1$ and $R_2 (> R_1)$. The surface charge on sphere 1 and 2 is $Q_1$ and $Q_2$, respectively and the spheres are surrounded everywhere by an electrolyte with Debye length $\lambda_D$.

The potential at $Z > R_2$ is the sum of the potential from sphere 1 and sphere 2. Based on Eq.7 the potential from sphere 1 is $V_1(Z) = \frac{k_e \cdot Q_1 \cdot \lambda_D}{\varepsilon \cdot Z \cdot R_1} \cdot e^{-\frac{Z}{\lambda_D}} \cdot \sinh\left(\frac{R_1}{\lambda_D}\right)$

And the potential from sphere 2 is $V_2(Z) = \frac{k_e \cdot Q_2 \cdot \lambda_D}{\varepsilon \cdot Z \cdot R_2} \cdot e^{-\frac{Z}{\lambda_D}} \cdot \sinh\left(\frac{R_2}{\lambda_D}\right)$.

Thus in the case of $Z > R_2$ the total potential is

$$V(Z) = \frac{k_e \cdot \lambda_D}{\varepsilon \cdot Z} \cdot e^{-\frac{Z}{\lambda_D}} \cdot \left\{ \frac{Q_1}{R_1} \cdot sinh\left(\frac{R_1}{\lambda_D}\right) + \frac{Q_2}{R_2} \cdot sinh\left(\frac{R_2}{\lambda_D}\right) \right\}. \tag{9}$$

If $Z > R_1$ but $Z < R_2$ the potential from sphere 1 can be obtained from Eq.7, $V_1(Z) = \frac{k_e \cdot Q_1 \cdot \lambda_D}{\varepsilon \cdot Z \cdot R_1} \cdot e^{-\frac{Z}{\lambda_D}} \cdot sinh\left(\frac{R_1}{\lambda_D}\right)$, but the potential from sphere 2 can be obtained from an equation like Eq.8, $V_2(Z) = \frac{k_e \cdot Q_2 \cdot \lambda_D}{\varepsilon \cdot Z \cdot R_2} \cdot e^{-\frac{R_2}{\lambda_D}} \cdot sinh\left(\frac{Z}{\lambda_D}\right)$.

Thus in the case of $R_1 < Z < R_2$ the total potential is:

$$V(Z) = \frac{k_e \cdot \lambda_D}{\varepsilon \cdot Z} \cdot \left\{ \frac{Q_1}{R_1} \cdot e^{-\frac{Z}{\lambda_D}} \cdot sinh\left(\frac{R_1}{\lambda_D}\right) + \frac{Q_2}{R_2} \cdot e^{-\frac{R_2}{\lambda_D}} \cdot sinh\left(\frac{Z}{\lambda_D}\right) \right\}. \tag{10}$$

Finally, if $Z < R_1$ the potential from sphere 1 can be obtained from Eq.8, $V_1(Z) = \frac{k_e \cdot Q_1 \cdot \lambda_D}{\varepsilon \cdot Z \cdot R_1} \cdot e^{-\frac{R_1}{\lambda_D}} \cdot sinh\left(\frac{Z}{\lambda_D}\right)$, while the potential from sphere 2 can be obtained from an equation similar to Eq.8, $V_2(Z) = \frac{k_e \cdot Q_2 \cdot \lambda_D}{\varepsilon \cdot Z \cdot R_2} \cdot e^{-\frac{R_2}{\lambda_D}} \cdot sinh\left(\frac{Z}{\lambda_D}\right)$.

Thus in the case of $R_1 > Z$ the total potential is:

$$V(Z) = \frac{k_e \cdot \lambda_D}{\varepsilon \cdot Z} \cdot sinh\left(\frac{Z}{\lambda_D}\right) \left\{ \frac{Q_1}{R_1} \cdot e^{-\frac{R_1}{\lambda_D}} + \frac{Q_2}{R_2} \cdot e^{-\frac{R_2}{\lambda_D}} \right\}. \tag{11}$$

### 2.3 Potential within the membrane of the charged vesicle – at high electrolyte concentration

Here we calculate the potential within the membrane of the charged vesicle when the electrolyte concentration is so high that the respective Debye length $\lambda_D$ is much shorter than the membrane thickness, i.e. $R_2 - R_1$ (see Figure 2). For example, if the ion concentration of the electrolyte is $1000 mol \cdot m^{-3}$, then the Debye length is about $0.304 nm$ while the membrane thickness is about $10 nm$.

In this case the screening effect of the electrolyte is so strong that if only a part of the straight line between a membrane charge and the P1 point (see Figure 2) crosses the intra-vesicular electrolyte the potential from that charge reduces to close to zero. Thus at the intra-vesicular surface of the membrane only those charges contribute to the potential at point P1 where $\alpha < \alpha_1$. These charges are not screened at all because there is no electrolyte in the membrane.

However, at the extra-vesicular surface of the membrane only those charges contribute to the potential at point P1 where $\alpha < \alpha_1 + \alpha_2$ ($\alpha_1$ and $\alpha_2$ are defined in Figure 2).

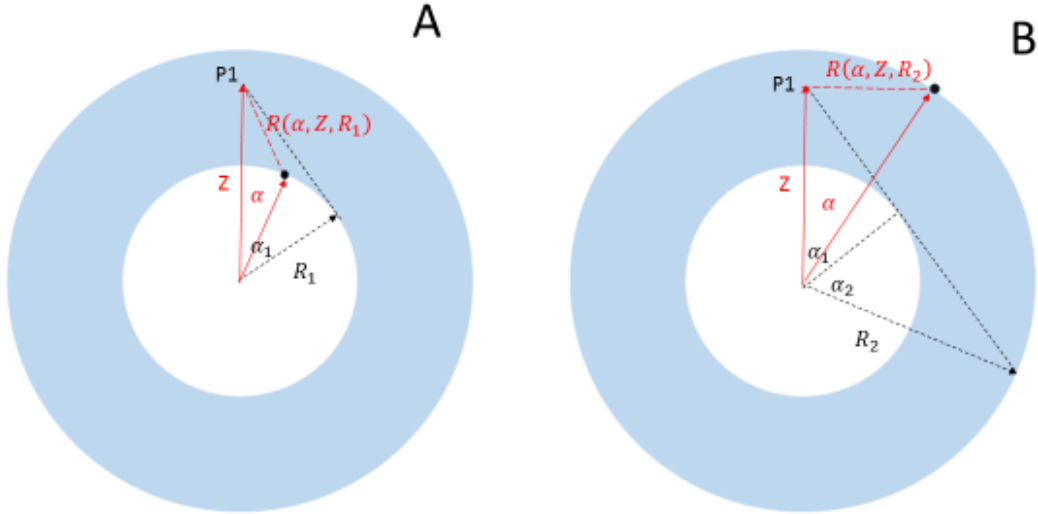

**Figure 2.** *Model of charged vesicle*
Light blue ring represents the membrane of the vesicle. $R_1$ and $R_2$ is the inner and outer radius of the vesicle membrane, respectively. P1: an intra-membrane point, at distance Z from the center of the vesicle, where we calculate the potential. A) In this figure we consider a charge (black dot) on the intra-vesicular surface of the membrane. $\alpha$ is the angle between the vectors pointing from the center of the vesicle to P1 and to the black dot. The line between P1 and the black dot is within the membrane (see red dashed line. Its length is marked by $R(\alpha, Z, R_1)$). When $\alpha$ is larger than $\alpha_1$ the line between P1 and the black dot is partly within the membrane and partly within the intra-vesicular electrolyte. B) In this figure we consider a charge (black dot) on the extra-vesicular surface of the membrane. $\alpha$ is the angle between the vectors pointing from the center of the vesicle to P1 and to the black dot. The line between P1 and the black dot is within the membrane. Its length is marked by $R(\alpha, Z, R_2)$. When $\alpha$ is larger than $\alpha_1 + \alpha_2$ the line between P1 and the black dot is partly within the membrane and partly within the intra-vesicular electrolyte.

At a point within the vesicle membrane, i.e. when $R_1 < Z < R_2$, the potential produced by the charges located at the intra-vesicular surface of the membrane is $V_1(Z)$ and by the charges located at the extra-vesicular surface of the membrane is $V_2(Z)$. Where (from Appendix 2)

$$V_1(Z) = \int_0^{\alpha_1} V_1(\alpha, Z) d\alpha = \frac{k_e \cdot Q_1}{2 \cdot \varepsilon_m} \int_0^{\alpha_1} \frac{\sin(\alpha) \cdot d\alpha}{R(\alpha, Z, R_1)} =$$

$$\frac{k_e \cdot Q_1}{2 \cdot \varepsilon_m R_1 \cdot Z} \left[ \sqrt{(Z-R_1)(Z+R_1)} - (Z-R_1) \right] \quad (12)$$

and

$$V_2(Z) = \int_0^{\alpha_1+\alpha_2} V_2(\alpha, Z) d\alpha = \frac{k_e \cdot Q_2}{2 \cdot \varepsilon_m} \int_0^{\alpha_1+\alpha_2} \frac{\sin(\alpha) \cdot d\alpha}{R(\alpha, Z, R_2)} =$$
$$\frac{k_e \cdot Q_2}{2 \cdot \varepsilon_m \cdot R_2 \cdot Z} \left[ \sqrt{Z^2 + R_2^2 - 2R_1^2 + 2\sqrt{(Z^2-R_1^2)(R_2^2-R_1^2)}} - (R_2 - Z) \right] \quad (13)$$

where $Q_1$ and $Q_2$ is the total charge of the intra-vesicular and extra-vesicular surface of the membrane, respectively, $\varepsilon_m = 2$ is the dielectric constant of the membrane. $R(\alpha, Z, R_1)$ and $R(\alpha, Z, R_2)$ are defined in the legends to Figure 2 (and can be calculated by Eq.5)

### 2.4 Potential within and around the charged vesicle membrane – at any electrolyte concentration

When the electrolyte concentration is not high enough then every charge of the vesicle membrane contributes to the potential. In this case the potential at a distance $Z$ from the center of the vesicle can be calculated by the following integrals:

$$V(Z) = \frac{k_e Q_1}{2} \int_0^\pi \frac{\sin(\alpha) d\alpha}{\varepsilon(Z) R(\alpha, Z, R_1)} e^{-\frac{R_{(e1)}(\alpha, Z)}{\lambda_D}} +$$
$$\frac{k_e Q_2}{2} \int_0^\pi \frac{\sin(\alpha) d\alpha}{\varepsilon(Z) R(\alpha, Z, R_2)} e^{-R_{(e2)}(\alpha, Z)/\lambda_D} \quad (14)$$

where $\lambda_D$ is the Debye length in the electrolyte and $R(\alpha, Z, R_1)$ is the distance between point P1 (located at a distance $Z$ from the center of the vesicle) and the charge located at the intra-vesicular surface of the membrane. The part of the distance that goes through electrolyte is marked by $R_{(e1)}(\alpha, Z)$. While $R(\alpha, Z, R_2)$ is the distance between point P1 and the charge located at the extra-vesicular surface of the membrane. The part of the distance that goes through electrolyte is marked by $R_{(e2)}(\alpha, Z)$. The dielectric constant is $\varepsilon(Z) = \varepsilon_m = 2$ at $R_1 < Z < R_2$ and $\varepsilon(Z) = \varepsilon_e = 80$ at $0 < Z < R_1$ and at $R_2 < Z$.

The integrals in Eq.14 can be calculated only numerically. Depending on the value of $Z$, the location of the charge and the value of angle $\alpha$ there are eleven cases for calculating $R_{(e1)}(\alpha, Z)$ and $R_{(e2)}(\alpha, Z)$ listed in Table 2.

**Table 2.** *Eleven cases for calculating $R_{(e1)}(\alpha, Z)$ and $R_{(e2)}(\alpha, Z)$*

| | Figures showing the eleven different cases | Equations for calculating $R_{(e1)}$ or $R_{(e2)}$ | Limits of angle $\alpha$ |
|---|---|---|---|
| I $(Z > R_2)$ Considered charge is on the outer sphere | 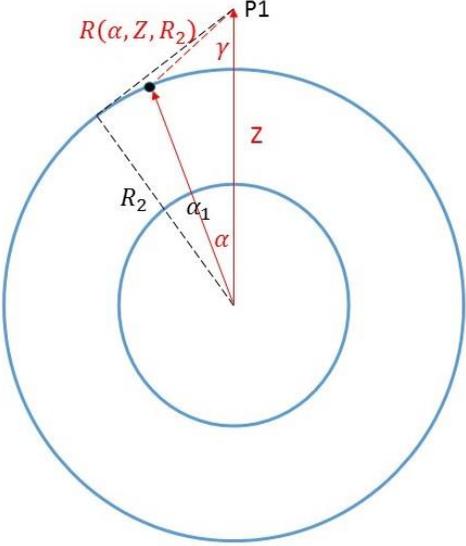 | $R_{(e2)}(\alpha, Z)$ $= R(\alpha, Z, R_2)$ | $0 < \alpha < \alpha_1$ $\cos(\alpha_1) = R_2/Z$ |
| II $(Z > R_2)$ Considered charge is on the outer sphere | 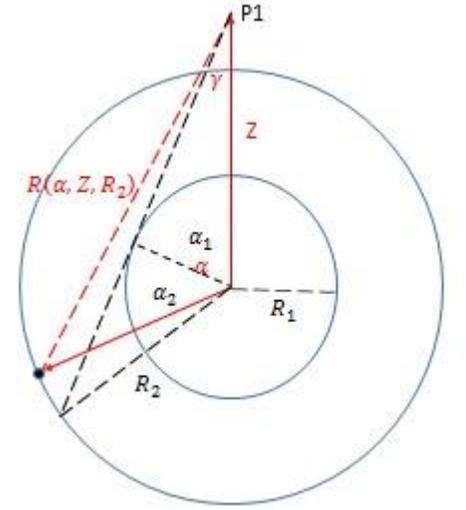 | $R_{(e2)}(\alpha, Z)$ $= R(\alpha, Z, R_2)$ $- r_2$ $r_2$ is the length of the chord (at the intersections of the outer circle and the red dashed line) and calculated by Eq.A13 | $\alpha_1 < \alpha$ $< \alpha_1 + \alpha_2$ $\cos(\alpha_1 + \alpha_2)$ is at Eq.A8 |

| | | | | |
|---|---|---|---|---|
| III<br><br>($Z > R_2$)<br>Considered charge is on the outer sphere | 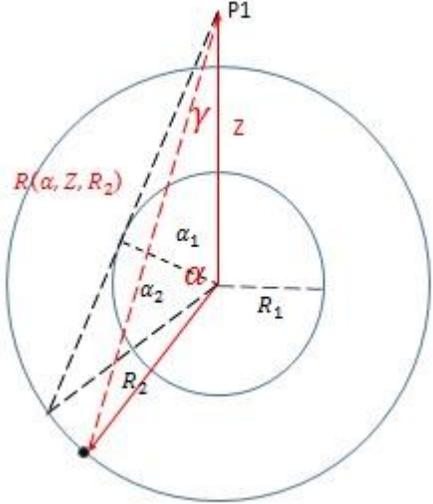 | $R_{(e2)}(\alpha, Z)$<br>$= R(\alpha, Z, R_2)$<br>$- (r_2 - r_1)$<br>$r_1$ is the length of the chord (at the intersections of the inner circle and the red dashed line) and calculated by Eq.A15 | $\alpha_1 + \alpha_2 < \alpha < \pi$<br><br>$\cos(\alpha_1 + \alpha_2)$ is at Eq.A8 |
| IV<br><br>($Z > R_2$)<br>Considered charge is on the inner sphere | 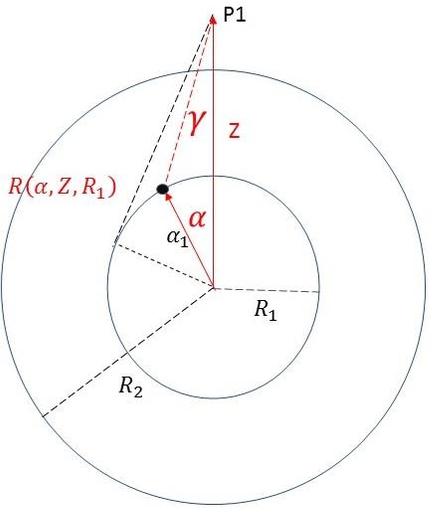 | $R_{(e1)}(\alpha, Z)$<br>$= R(\alpha, Z, R_1)$<br>$- 0.5 \cdot (r_2 - r_1)$ | $0 < \alpha < \alpha_1$<br><br>$\cos(\alpha_1) = R_1/Z$ |
| V<br><br>($Z > R_2$)<br>Considered charge is on the inner sphere | 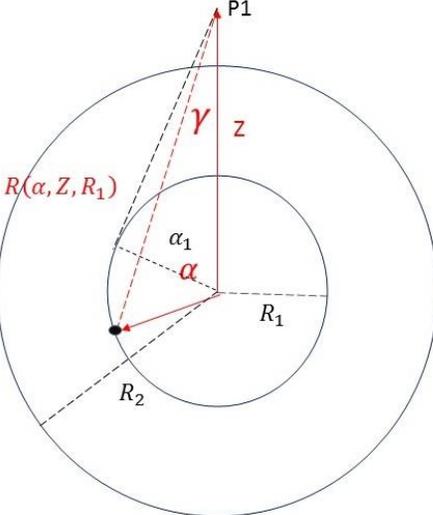 | $R_{(e1)}(\alpha, Z)$<br>$= R(\alpha, Z, R_1)$<br>$- 0.5 \cdot (r_2 - r_1)$ | $\alpha_1 < \alpha < \pi$<br><br>$\cos(\alpha_1) = R_1/Z$ |
| VI<br><br>($R_1 < Z < R_2$)<br>Considered charge is on the outer sphere | 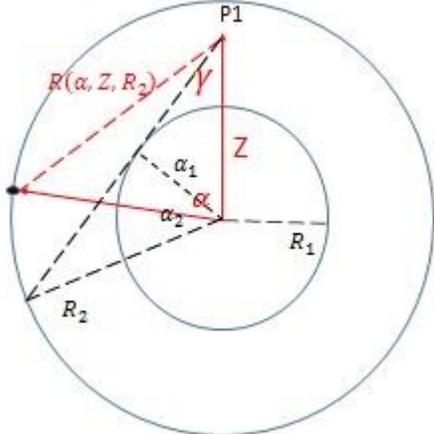 | $R_{(e2)}(\alpha, Z) = 0$ | $0 < \alpha < \alpha_1 + \alpha_2$<br><br>$\cos(\alpha_1 + \alpha_2)$ is at Eq.A8 |
| VII<br><br>($R_1 < Z < R_2$)<br>Considered charge is on the outer sphere | 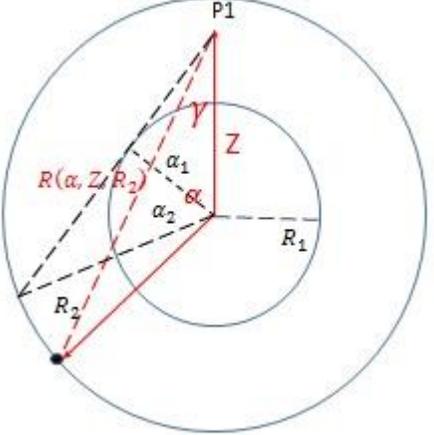 | $R_{(e2)}(\alpha, Z) = r_1$ | $\alpha_1 + \alpha_2 < \alpha < \pi$<br><br>$\cos(\alpha_1 + \alpha_2)$ is at Eq.A8 |
| VIII<br><br>($R_1 < Z < R_2$)<br>Considered charge is on the inner sphere | 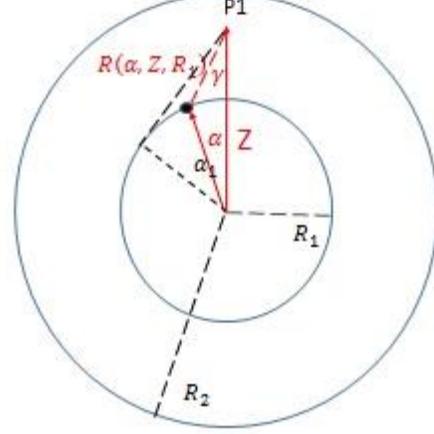 | $R_{(e1)}(\alpha, Z) = 0$ | $0 < \alpha < \alpha_1$<br><br>$\cos(\alpha_1) = R_1/Z$ |
| IX<br><br>($R_1 < Z < R_2$)<br>Considered charge is on the inner sphere | | $R_{(e1)}(\alpha, Z) = r_1$ | $\alpha_1 < \alpha < \pi$<br><br>$\cos(\alpha_1) = R_1/Z$ |
| X<br><br>($Z < R_1$)<br>Considered charge is on the outer sphere | | $R_{(e2)}(\alpha, Z)$<br>$= R(\alpha, Z, R_2)$<br>$- 0.5 \cdot (r_2 - r_1)$ | $0 < \alpha < \pi$ |
| XI<br><br>($Z < R_1$)<br>Considered charge is on the inner sphere | | $R_{(e1)}(\alpha, Z) = 0$ | $0 < \alpha < \pi$ |

### 3. Results

The potential, $V$, inside and around a charged sphere of radius $R_1$ is calculated by Eqs.7,8. The radius of the charged sphere, $R_1 = 10^{-7}m = 100nm$, is a typical size of a large unilamellar vesicle (LUV) [8]. The total charge of the sphere is: $Q_1 = -3.35 \cdot 10^{-14}C$. This is the total charge of a PLFE (bipolar tetraether lipid with the polar lipid fraction E) vesicle of radius $R_1$ if the cross sectional area of a PLFE is $0.6 nm^2$ and the charge of a PLFE molecule is $1.6 \cdot 10^{-19}C$ [1,2]. In Figure 3 the potential multiplied by $-1$ is plotted as a function of the radial distance, $Z$, from the center of the sphere in the case of different monovalent ion concentrations of the electrolyte.

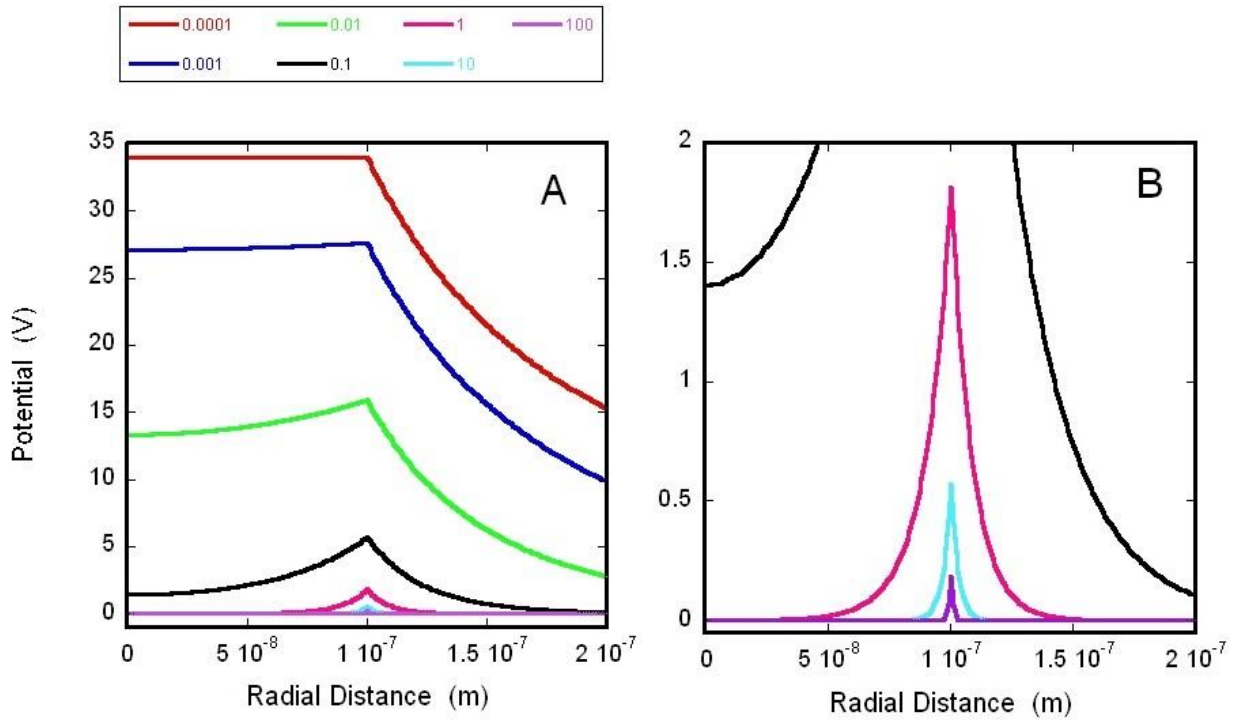

**Figure 3.** *Calculated potential inside and around a charged sphere*
   A) The potential, $-V$, as a function of the radial distance from the center of the charged sphere is calculated by Eqs.7,8. The radius of the charged sphereis: $R_1 = 10^{-7}m = 100nm$. The total charge of the sphere is: $Q_1 = -3.35 \cdot 10^{-14}C$. Each curve was calculated at a certain monovalent ion concentration of the electrolyte. The concentrations belonging to a certain color are shown in $mol/m^3$ above Figure 3A. The Debye lengths (in $m$) belonging to each concentration are listed in Table 1.
   B) Calculated potentials within the region: $0V < -V(Z) < 2V$.

**Table 1.** *Monovalent ion concentrations of electrolytes and the respective Debye lengths*

| Electrolyte concentration ($mol \cdot m^{-3}$) | Debye length, $\lambda_D$ (m) |
|---|---|
| 0.0001 | $9.62 \cdot 10^{-7}$ |
| 0.001 | $3.04 \cdot 10^{-7}$ |
| 0.01 | $9.62 \cdot 10^{-8}$ |
| 0.1 | $3.04 \cdot 10^{-8}$ |
| 1.0 | $9.62 \cdot 10^{-9}$ |
| 10.0 | $3.04 \cdot 10^{-9}$ |
| 100.0 | $9.62 \cdot 10^{-10}$ |

*Note, that the dielectric constant of the electrolyte decreases with increasing ion concentration. However in the above concentration region the decrease is within one percent [9,10]. Thus in our calculations the dielectric constant is taken $\varepsilon_e = 80$ at the above electrolyte concentrations.

By using Eqs.9-11 the potential was calculated within and around two concentric charged spheres (see Figure 4).

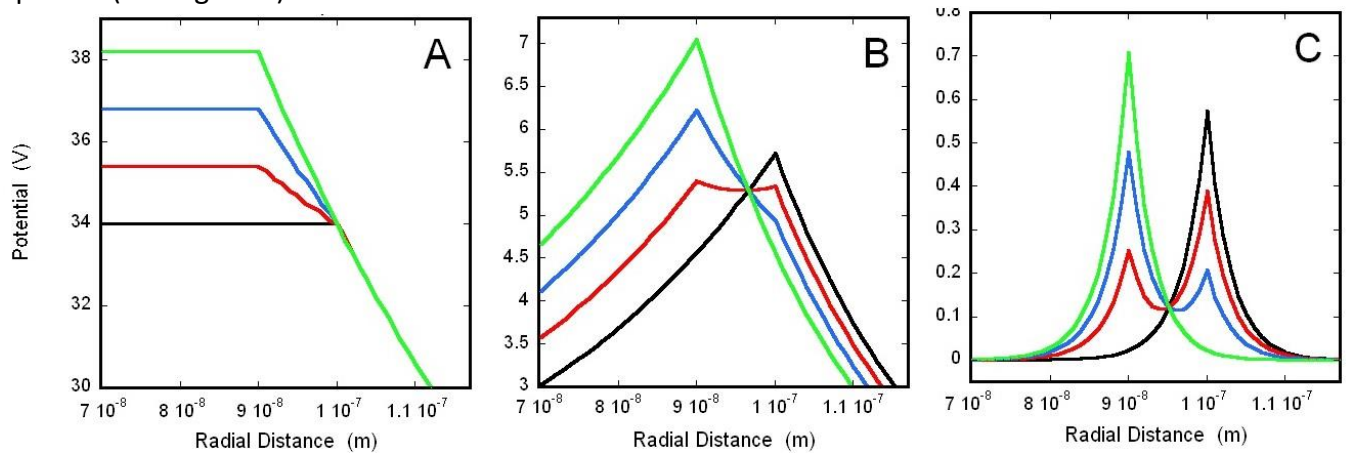

**Figure 4.** *Potential within and around two concentric charged sphere surrounded by electrolyte*
The potential, $-V$, as a function of the radial distance from the center of the charged spheres is calculated by Eqs.9-11.
The radius of the smaller and larger sphere is $R_1 = 0.9 \cdot 10^{-7} m$ and $R_2 = 10^{-7} m$, respectively, while the total charge of the smaller and larger sphere is $Q_1$ and $Q_2$. In our calculations always $Q_1 + Q_2 = Q = -3.35 \cdot 10^{-14} C$. In the case of black line: $Q_2 = Q$, red line: $Q_2 = 2Q/3$, blue line: $Q_2 = Q/3$, green line: $Q_2 = 0C$. A) Concentration of electrolyte (of monovalent ions) is $0.0001 \, mol/m^3$ and the respective Debye length is $\lambda_D = 9.62 \cdot 10^{-7} m$. B) Concentration of electrolyte (of monovalent ions) is $0.1 \, mol/m^3$ and the respective Debye length is $\lambda_D = 3.04 \cdot 10^{-8} m$. C) Concentration of electrolyte (of monovalent ions) is $10 \, mol/m^3$ and the respective Debye length is $\lambda_D = 3.04 \cdot 10^{-9} m$.

The sum of Eq.12 and Eq.13 gives the potential within the membrane of the vesicle filled and surrounded by electrolyte with high ion concentration.

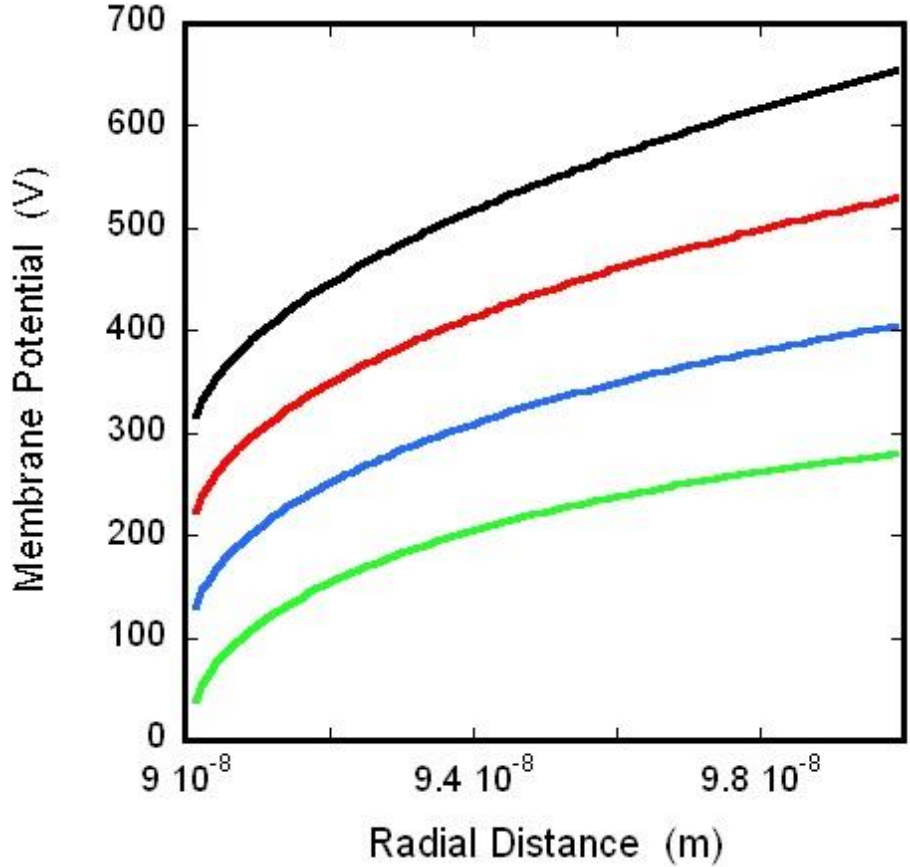

**Figure 5.** *Potential within the vesicle membrane at high electrolyte concentration*
The vesicle membrane potential, $-V$, as a function of the radial distance from the center of the vesicle is calculated by the sum of Eq.12 and Eq.13. These equations are valid only at high ion concentrations at the intra- and extra-vesicular electrolyte, i.e.: when $\lambda_D \ll (R_2 - R_1)$. The radius of the inner- and outer surface of the vesicle membrane is $R_1 = 0.9 \cdot 10^{-7} m$ and $R_2 = 10^{-7} m$, respectively, while the total charge of the inner and outer surface of the vesicle membrane is $Q_1$ and $Q_2$. In our calculations always $Q_1 + Q_2 = Q = -3.35 \cdot 10^{-14} C$. In the case of the black curve: $Q_2 = Q$, red curve: $Q_2 = 2Q/3$, blue curve: $Q_2 = Q/3$, green curve: $Q_2 = 0C$. The dielectric constant of the membrane is $\varepsilon_m = 2$.

In Eq.14 after performing the numerically integrations we get the radial dependence of the potential of charged vesicle (see Figure 6).

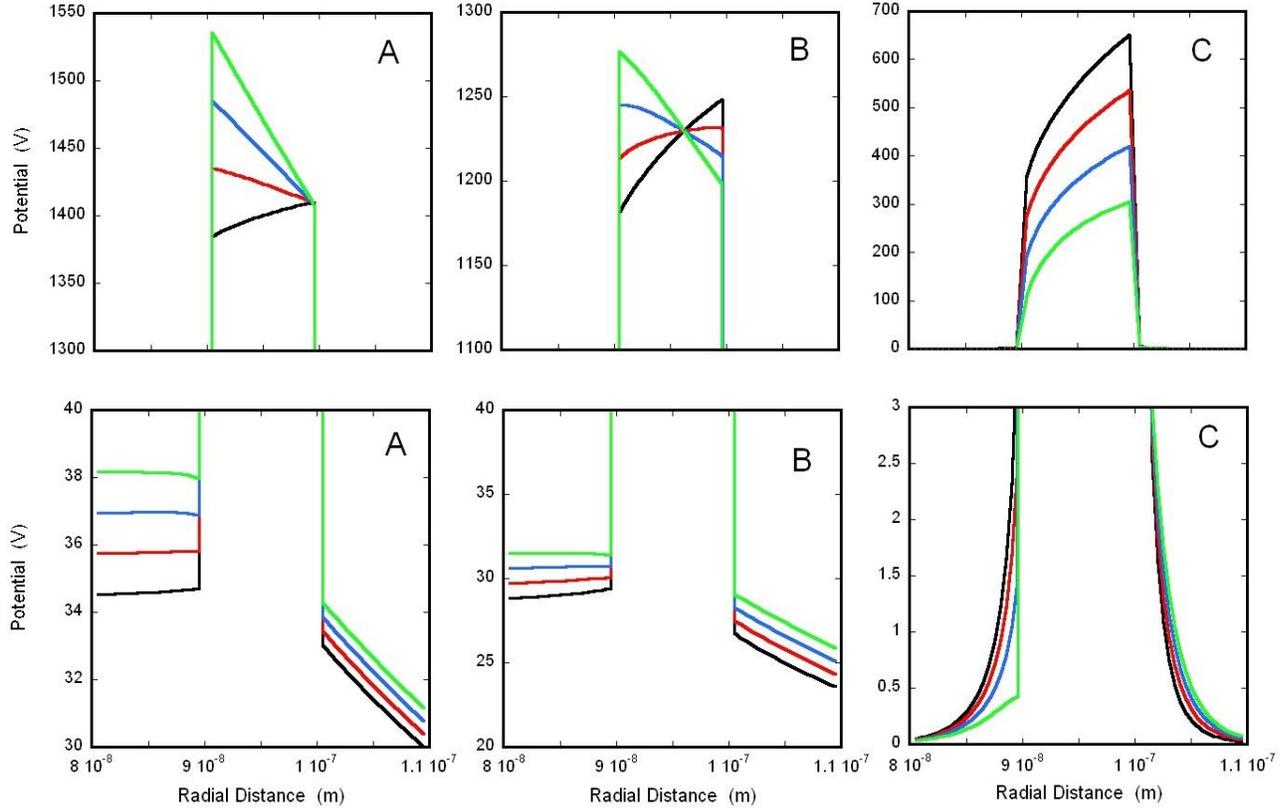

**Figure 6.** *Potential within and around the charged vesicle membrane at different electrolyte concentrations*

The potential, $-V$, within and around the charged vesicle membrane is calculated by Eq.14. The radius of the inner- and outer surface of the vesicle membrane is $R_1 = 0.9 \cdot 10^{-7}m$ and $R_2 = 10^{-7}m$, respectively, while the total charge of the inner and outer surface of the vesicle membrane is $Q_1$ and $Q_2$. In our calculations always $Q_1 + Q_2 = Q = -3.35 \cdot 10^{-14}C$. In the case of the black curve: $Q_2 = Q$, red curve: $Q_2 = 2Q/3$, blue curve: $Q_2 = Q/3$, green curve: $Q_2 = 0C$. A) Concentration of electrolyte (of monovalent ions) is $0.0001\ mol/m^3$ and the respective Debye length is $\lambda_D = 9.62 \cdot 10^{-7}m$. B) Concentration of electrolyte (of monovalent ions) is $0.001\ mol/m^3$ and the respective Debye length is $\lambda_D = 3.04 \cdot 10^{-7}m$. C) Concentration of electrolyte (of monovalent ions) is $10\ mol/m^3$ and the respective Debye length is $\lambda_D = 3.04 \cdot 10^{-9}m$. Note, each subfigure is divided into a high- and low-potential part.

## 4. Discussion

By using Eqs.7,8 one can calculate the potential of a charged sphere filled and surrounded by electrolyte. These equations are generalization of the Shell Theorem (given by Eqs.1,2) where the charged sphere is in vacuum. One can get from Eqs.7,8 the equations of the Shell Theorem by taking infinite long Debye length (that is characteristic for vacuum):

at $Z > R_1$

$$V(Z) = \lim_{\lambda_D \to \infty} \left\{ \frac{k_e \cdot Q_1 \cdot \lambda_D}{\varepsilon \cdot Z \cdot R_1} \cdot e^{-\frac{Z}{\lambda_D}} \cdot sinh\left(\frac{R_1}{\lambda_D}\right) \right\} =$$
$$\lim_{\lambda_D \to \infty} \left\{ \frac{k_e \cdot Q_1 \cdot \lambda_D}{\varepsilon \cdot Z \cdot R_1} \cdot e^{-\frac{Z}{\lambda_D}} \cdot \left[\frac{R_1}{\lambda_D} + \frac{1}{3!}\left(\frac{R_1}{\lambda_D}\right)^3 + \frac{1}{5!}\left(\frac{R_1}{\lambda_D}\right)^5 + \cdots \right] \right\} = \frac{k_e \cdot Q_1}{\varepsilon \cdot Z} \quad (15)$$

and at $Z < R_1$

$$V(Z) = \lim_{\lambda_D \to \infty} \left\{ \frac{k_e \cdot Q_1 \cdot \lambda_D}{\varepsilon \cdot Z \cdot R_1} \cdot e^{-\frac{R_1}{\lambda_D}} \cdot sinh\left(\frac{Z}{\lambda_D}\right) \right\} =$$
$$\lim_{\lambda_D \to \infty} \left\{ \frac{k_e \cdot Q_1 \cdot \lambda_D}{\varepsilon \cdot Z \cdot R_1} \cdot e^{-\frac{R_2}{\lambda_D}} \cdot \left[\frac{Z}{\lambda_D} + \frac{1}{3!}\left(\frac{Z}{\lambda_D}\right)^3 + \frac{1}{5!}\left(\frac{Z}{\lambda_D}\right)^5 + \cdots \right] \right\} = \frac{k_e \cdot Q_1}{\varepsilon \cdot R_1}. \quad (16)$$

According to the Shell Theorem the potential inside the charged sphere is constant. Similarly, at low ion concentrations (between $0 - 0.001 mol/m^3$), the potential close to constant inside the charged sphere (see Figure 3A). At higher ion concentrations however as a consequence of the increased screening the potential drops both outward and inward from the surface of the charged sphere (see Figure 3B).

Similar to the result of the Shell Theorem In the case of two concentric charged spheres, surrounded by electrolyte of low ion concentration, the potential is constant within the smaller sphere (see Figure 4A). If there is any amount of charge on the inner sphere the potential linearly decreases from the inner to the outer sphere. While the potential is constant from the center to the outer sphere if only the outer sphere is charged. At high ion concentrations, because of the high screening, however, the potential has two maxima at the radii of the charged spheres (see Figure 4C).

In the case of charged vesicles the charges are located at the inner and outer surface of the vesicle membrane. Like in the case of the above mentioned charged double spheres the intra- and extra-vesicular space is filled by electrolyte, but there is no electrolyte between the two charged spheres. The inner part of the lipid membrane is hydrophobic. If the electrolyte concentration is high the membrane potential can be calculated analytically by Eqs.12,13. Figure 5 shows the calculated membrane potential if the total membrane charge is divided between the outer and inner membrane surface on four different ways. Interestingly, in each case the membrane potential increases with increasing radial distance even if the charge is zero at the outer membrane surface (see green line in Figure 5). As we mentioned before in this case the screening effect of the electrolyte is so strong that if only a part of the straight line between a membrane charge and the P1 point (see Figure 2) crosses the intra-vesicular electrolyte the potential from that charge reduces to zero. Thus at the intra-vesicular surface of the membrane only those charges contribute to the potential at point P1 where $\alpha < \alpha_1$. However, at the extra-vesicular surface of the membrane only those charges contribute to the potential at point P1 where $\alpha < \alpha_1 + \alpha_2$ ($\alpha_1$ and $\alpha_2$ are defined in Figure 2). The charges contributing to the potential at point P1 are situated on two spherical caps. The surface area of the spherical cap on the intra-vesicular surface of the membrane is:

$$S_1 = 2 \cdot R_1^2 \cdot \pi \cdot [1 - \cos(\alpha_1)] = 2 \cdot R_1^2 \cdot \pi \cdot \left[1 - \frac{R_1}{Z}\right] \qquad (17)$$

while based on Eq.A8 on the extra-vesicular surface of the membrane the surface area of the spherical cap is:

$$S_2 = 2 \cdot R_2^2 \cdot \pi \cdot [1 - \cos(\alpha_1 + \alpha_2)] = 2 \cdot R_2^2 \cdot \pi \cdot \left[1 - \frac{R_1^2 - \sqrt{(Z^2 - R_1^2)(R_2^2 - R_1^2)}}{Z \cdot R_2}\right] \qquad (18)$$

Thus the total charge contributing to the potential at point P1 is:

$$\rho_1 \cdot S_1 + \rho_2 \cdot S_2 = \frac{Q_1}{2}\left[1 - \frac{R_1}{Z}\right] + \frac{Q_2}{2}\left[1 - \frac{R_1^2 - \sqrt{(Z^2 - R_1^2)(R_2^2 - R_1^2)}}{Z \cdot R_2}\right] \qquad (19)$$

In Figure 7 $(\rho_1 \cdot S_1 + \rho_2 \cdot S_2)/Q$ is plotted against the radial distance of point P1

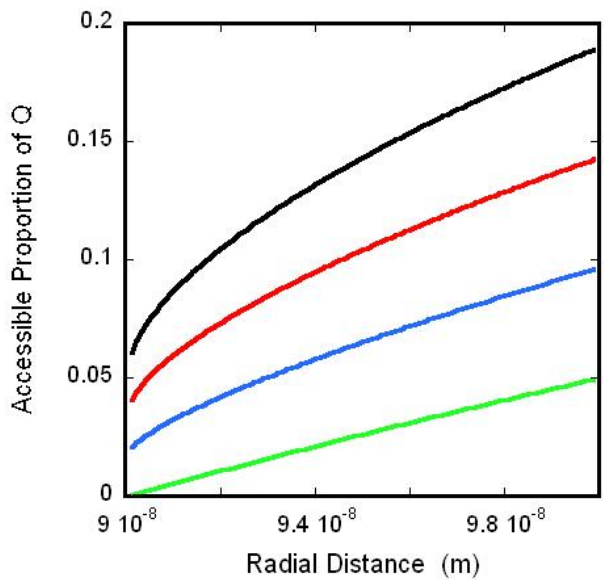

**Figure 7.** *Proportion of total charge contributing to the potential within the vesicle membrane*
To calculate the accessible proportion of the total charge of the vesicle membrane we used Eq.19. The radius of the inner- and outer surface of the vesicle membrane is $R_1 = 0.9 \cdot 10^{-7} m$ and $R_2 = 10^{-7} m$, respectively, while the total charge of the inner and outer surface of the vesicle membrane is $Q_1$ and $Q_2$. In our calculations always $Q_1 + Q_2 = Q = -3.35 \cdot 10^{-14} C$. In the case of the black curve: $Q_2 = Q$, red curve: $Q_2 = 2Q/3$, blue curve: $Q_2 = Q/3$, green curve: $Q_2 = 0C$.

The similar characteristics of the curves in Figure 5 and Figure 7 shows that the potential in the vesicle membrane close to proportional to the amount of charges that are not screened out by the highly concentrated electrolyte surrounding the vesicle membrane. Note also that the membrane potential calculated from the sum of Eq.12 and Eq.13 is similar to the result of the numerical integration of Eq.14 (compare Figure 5 with Figure 6C where the electrolyte concentration is $10 \ mol/m^3$).
In the case of low electrolyte concentration, when every charge of the vesicle membrane contributes to the potential, the potential close to linearly decreases within the membrane with increasing radial distance (see Figure 6A). This is similar to the change of the potential between two charged spheres immersed into electrolyte with low ion concentration (see Figure 4A). This can be explained by the Shell Theorem that works well at low electrolyte concentration. According to Eq.2 the membrane potential, caused by the charges on the intra-vesicular surface, is close to linearly decreasing within the interval $(R_1, R_2)$ because $\frac{[R_2 - R_1]}{R_2} \ll 1$. On the other hand, according to Eq.1, the membrane potential, created by the charges on the extra-vesicular surface of the membrane, is constant. Thus the total membrane potential decreases also close to linearly. When there is no charge on the intra-vesicular surface, i.e. $Q_2 = Q$, based on Eq.1 one may expect constant membrane potential. However, the membrane potential is not constant but slightly increasing (see black curve in Figure 6A). This maybe the case because the ion concentration of the electrolyte is slightly higher than zero. With further increasing electrolyte concentration the membrane potential belonging to $Q_2 = 2 \cdot Q/3$ increases too (see red curve in Figure 6B). Finally we should mention the fundamental differences between Figure 4A and Figure 6A. Since the dielectric constant in the membrane is 40 times smaller than in the intra- and extra-vesicular space there is a sudden 40 times jump and 40 times drop of the potential at $R_1$ and $R_2$, respectively.

## Conclusions
By using the Debye screened potential a generalized version of the Shell Theorem is developed and analytical equations are derived to calculate i) the potential of a charged sphere surrounded by electrolyte, ii) the potential of two concentric charged spheres surrounded by electrolyte, iii) the membrane potential of a charged lipid vesicle surrounded by electrolyte with high ion concentration. By numerical integration the potential of a lipid vesicle is calculated at any electrolyte concentration. In general with increasing ion concentration the screening effect of the electrolyte increases and the overall potential of the above mentioned spheres, vesicles is decreasing. With increasing distance from the outer surface of the charged sphere or vesicle the potential decrease is steeper in the case of higher electrolyte concentration. Inside the charged sphere or vesicle at low electrolyte concentration the potential is close to constant. However with increasing electrolyte concentration the decreases of the potential towards the center of the sphere (or vesicle) becomes steeper.

*Appendix 1*
Let us do the following substitution in the integral (in Eq.6): $u = \cos(\alpha)$.
Thus in Eq.6 $\sin(\alpha)\, d\alpha$ can be substituted by $-du$ and we get

$$V(Z) = \frac{k_e \cdot \rho_1 \cdot 2R_1^2 \pi}{\varepsilon} \int_{-1}^{1} \frac{1}{\sqrt{R_1^2 + Z^2 - 2ZR_1 u}} e^{-\sqrt{R_1^2 + Z^2 - 2ZR_1 u}/\lambda_D} du \quad \text{(A1)}$$

Finally, let us do this substitution in Eq.A1: $w = -\sqrt{R_1^2 + Z^2 - 2ZR_1 u}/\lambda_D$ and thus

$$dw = \frac{ZR_1}{\lambda_D \sqrt{R_1^2 + Z^2 - 2ZR_1 u}} du \text{ and we get}$$

$$V(Z) = \frac{k_e \cdot \rho_1 \cdot 2R_1^2 \pi \cdot \lambda_D}{\varepsilon \cdot Z \cdot R_1} \int_{w(u=-1)}^{w(u=1)} e^w \, dw = \frac{k_e \rho_1 2R_1^2 \pi \lambda_D}{\varepsilon Z R_1} [e^w]_{w(u=-1)}^{w(u=1)} \quad \text{(A2)}$$

where

$$w(u=-1) = -\frac{\sqrt{R_1^2 + Z^2 + 2ZR_1}}{\lambda_D} = -\frac{(Z+R_1)}{\lambda_D}, \text{ while}$$

in the case of $Z > R_1$

$$w(u=1) = -\frac{\sqrt{R_1^2 + Z^2 - 2ZR_1}}{\lambda_D} = -\frac{\sqrt{(Z-R_1)^2}}{\lambda_D} = -\frac{(Z-R_1)}{\lambda_D}, \text{ and}$$

in the case of $Z < R_1$

$$w(u=1) = -\frac{\sqrt{R_1^2 + Z^2 - 2ZR_1}}{\lambda_D} = -\frac{\sqrt{(R_1-Z)^2}}{\lambda_D} = -\frac{(R_1-Z)}{\lambda_D}.$$

Here we mention that the square root of a positive number, $(Z-R_1)^2$ or $(R_1-Z)^2$ should be positive (only imaginary number's square root is negative). Thus the square root is $Z - R_1$ if $Z > R_1$ and the square root is $R_1 - Z$ if $Z < R_1$.

Thus the potential at $Z > R_1$ is

$$V(Z) = \frac{k_e \cdot \rho_1 \cdot 2R_1^2 \pi \cdot \lambda_D}{\varepsilon \cdot Z \cdot R_1} \left[ e^{-\frac{(Z-R_1)}{\lambda_D}} - e^{-\frac{(Z+R_1)}{\lambda_D}} \right] = \frac{k_e \cdot Q_1 \cdot \lambda_D}{\varepsilon \cdot Z \cdot R_1} \cdot e^{-\frac{Z}{\lambda_D}} \cdot \sinh\left(\frac{R_1}{\lambda_D}\right) \quad \text{(A3)}$$

Where $Q_1 = 4R_1^2 \pi \cdot \rho_1$ is the total charge of the sphere.
Finally, the potential at $Z < R_1$ is

$$V(Z) = \frac{k_e \cdot \rho_1 \cdot 2R_1^2 \pi \cdot \lambda_D}{\varepsilon \cdot Z \cdot R_1} \left[ e^{-\frac{(R_1-Z)}{\lambda_D}} - e^{-\frac{(Z+R_1)}{\lambda_D}} \right] = \frac{k_e \cdot Q_1 \cdot \lambda_D}{\varepsilon \cdot Z \cdot R_1} \cdot e^{-\frac{R_1}{\lambda_D}} \cdot \sinh\left(\frac{Z}{\lambda_D}\right) \quad \text{(A4)}$$

*Appendix 2*
First we calculate the potential within the membrane induced by the charges located at the intra-vesicular membrane surface, $V_1(Z)$ (see Figure 2A) where $R_1 < Z < R_2$:

$$V_1(Z) = \int_0^{\alpha_1} V_1(\alpha, Z) d\alpha = \int_0^{\alpha_1} \frac{k_e \cdot \rho_1 \cdot 2 \cdot R_1 \cdot \sin(\alpha) \cdot \pi \cdot R_1 \cdot d\alpha}{\varepsilon_m R(\alpha, Z, R_1)} \quad (A5)$$

Note, that at $\alpha > \alpha_1$ the line between point P1 and any one charge of the intra-vesicular membrane surface partially goes through the intra-vesicular electrolyte. The electrolyte's ion concentration is assumed to be so high that the screening reduces the potential of any one of those charges to zero. $\alpha_1$ is defined in Figure 2A and its cosine is: $\cos(\alpha_1) = R_1/Z$. However at $0 < \alpha < \alpha_1$ the entire line between any of those charges and point P1 is in the membrane and the potential is not screened at all (see Eq.A5). After substituting Eq.5 into Eq.A5 and doing substitution $u = \cos(\alpha)$ we get:

$$V_1(Z) = \frac{k_e \cdot \rho_1 \cdot 2R_1^2 \pi}{\varepsilon_m} \int_{R_1/Z}^{1} \frac{1}{\sqrt{R_1^2 + Z^2 - 2ZR_1 u}} du = \frac{k_e \cdot \rho_1 \cdot 2R_1^2 \pi}{\varepsilon_m} \cdot \frac{\left[\sqrt{R_1^2 + Z^2 - 2ZR_1 u}\right]_{R_1/Z}^{1}}{-R_1 Z} =$$

$$\frac{k_e \cdot Q_1}{2 \cdot \varepsilon_m \cdot R_1 \cdot Z} \cdot \left[\sqrt{(Z-R_1)(Z+R_1)} - (Z-R_1)\right] \quad (A6)$$

$V_2(Z)$ is the potential at point P1 generated by the charges located at the extra-vesicular surface of the membrane. In this case charges located within $0 < \alpha < \alpha_1 + \alpha_2$ generate potential without any screening, while charges located at $\alpha > \alpha_1 + \alpha_2$ are screened out completely. Note that $\alpha_2$ is defined in Figure 2B and its cosine is $\cos(\alpha_2) = R_1/R_2$. Thus

$$V_2(Z) = \int_0^{\alpha_1 + \alpha_2} \frac{k_e \cdot \rho_2 \cdot 2 \cdot R_2 \cdot \sin(\alpha) \cdot \pi \cdot R_2 \cdot d\alpha}{\varepsilon_m R(\alpha, Z, R_2)} =$$

$$\frac{k_e \cdot Q_2}{2 \cdot \varepsilon_m} \int_{\cos(\alpha_1 + \alpha_2)}^{1} \frac{1}{\sqrt{R_2^2 + Z^2 - 2ZR_2 u}} du = \frac{k_e \cdot Q_2}{2 \cdot \varepsilon_m} \cdot \frac{\left[\sqrt{R_2^2 + Z^2 - 2ZR_2 u}\right]_{\cos(\alpha_1 + \alpha_2)}^{1}}{-R_2 Z} =$$

$$\frac{k_e \cdot Q_2}{2 \cdot \varepsilon_m \cdot R_2 \cdot Z} \cdot \left[\sqrt{Z^2 + R_2^2 - 2R_1^2 + 2\sqrt{(Z^2 - R_1^2)(R_2^2 - R_1^2)}} - (R_2 - Z)\right] \quad (A7)$$

where we used

$$\cos(\alpha_1 + \alpha_2) = \cos(\alpha_1)\cos(\alpha_2) - \sin(\alpha_1)\sin(\alpha_2) =$$
$$\cos(\alpha_1)\cos(\alpha_2) - \sqrt{1 - \cos^2(\alpha_1)} \cdot \sqrt{1 - \cos^2(\alpha_2)} =$$

$$\frac{R_1}{Z} \cdot \frac{R_1}{R_2} - \sqrt{1 - (R_1/Z)^2} \cdot \sqrt{1 - (R_1/R_2)^2} = \frac{R_1^2 - \sqrt{(Z^2 - R_1^2)(R_2^2 - R_1^2)}}{Z \cdot R_2} \quad (A8)$$

*Appendix 3*

Here we calculate the length of the chord, $r_2$ shown in Figure A1. The equation of the red line is:

$$y = tg(\gamma) \cdot (x + Z) \tag{A9}$$

And the equation of the circle of radius $R_2$ is:

$$x^2 + y^2 = R_2^2 \tag{A10}$$

The solutions of Eqs.A9,A10 results in the coordinates of the two intersections between the straight line and the circle. The solutions are:

$$x_{2/1} = \frac{-Z \cdot tg^2(\gamma) \pm \sqrt{R_2^2 \cdot [tg^2(\gamma)+1] - tg^2(\gamma) \cdot Z^2}}{tg^2(\gamma)+1} \tag{A11}$$

$$y_{2/1} = tg(\gamma) \cdot [x_{2/1} + Z] \tag{A12}$$

From Eqs.A11,A12 we get the length of the chord:

$$r_2 = \sqrt{(x_2-x_1)^2 + (y_2-y_1)^2} = 2 \cdot \sqrt{\frac{R_2^2 \cdot [tg^2(\gamma)+1] - tg^2(\gamma) \cdot Z^2}{tg^2(\gamma)+1}} \tag{A13}$$

and based on Figure A1 $tg(\gamma)$ depends on $\alpha$ as follows:

$$tg(\gamma) = \frac{R_2 \cdot \sin(\alpha)}{Z - R_2 \cdot \cos(\alpha)} \tag{A14}$$

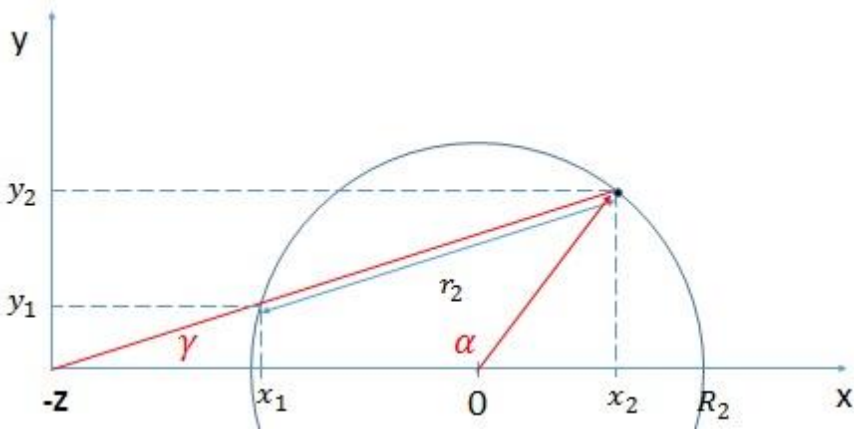

**Figure A1.** *Calculating the length of the chord, $r_2$*

If the straight line intersects the circle with radius $R_1$ the length of the chord, $r_1$ is:

$$r_1 = 2 \cdot \sqrt{\frac{R_1^2 \cdot [tg^2(\gamma)+1] - tg^2(\gamma) \cdot Z^2}{tg^2(\gamma)+1}} \tag{A15}$$

and $tg(\gamma)$ depends on $\alpha$ as follows:

$$tg(\gamma) = \frac{R_1 \cdot \sin(\alpha)}{Z - R_1 \cdot \cos(\alpha)}. \tag{A16}$$